\documentclass[prd,twocolumn,aps, superscriptaddress, floatfix,preprintnumbers,showpacs]{revtex4}
\usepackage{amsthm}   
\usepackage{amsmath}  
\usepackage{amssymb}  
\usepackage{mathrsfs}
\usepackage{color}
\usepackage[all]{xy}
\usepackage{graphicx}
\usepackage{indentfirst}

\newcommand{\pa}{\partial}

\newcommand{\be}{\begin{equation}}
\newcommand{\ee}{\end{equation}}
\newcommand{\ba}{\begin{eqnarray}}
\newcommand{\ea}{\end{eqnarray}}

\newcommand{\en}{\nonumber\\}

\newcommand{\de}{\delta}

\newcommand{\kk}{\mathbf{k}}
\newcommand{\xx}{\mathbf{x}}
\newcommand{\yy}{\mathbf{y}}
\newcommand{\zz}{\mathbf{z}}

\newcommand{\s}{\sigma}

\newcommand{\rhob}{\bar{\rho}}
\newcommand{\fnl}{f_{\rm NL}}

\begin{document}
\title{Oscillating Bispectra and Galaxy Clustering: A Novel Probe of Inflationary Physics with Large-Scale Structure}
\author{Francis-Yan Cyr-Racine}\email{francis@phas.ubc.ca}
\affiliation{Department of Physics and Astronomy, University of British Columbia, Vancouver, BC, V6T 1Z1, Canada}
\affiliation{California Institute of Technology, Pasadena, CA 91125, USA}
\author{Fabian Schmidt}\email{fabians@caltech.edu}
\affiliation{California Institute of Technology, Pasadena, CA 91125, USA}
\date{\today}
\begin{abstract}
Many models of inflation predict oscillatory features in the bispectrum of primordial fluctuations. Since it has been shown that primordial non-Gaussianity can lead to a scale-dependent halo bias, we investigate the effect of oscillations in the three-point function on the clustering of dark-matter halos. Interestingly, we find that features in the inflaton potential such as oscillations or sharp steps get imprinted in the mass dependence of the non-Gaussian halo bias. In this paper, we focus on models displaying a sharp feature in the inflaton potential as well as resonant non-Gaussianity. In both cases, we find a strong scale dependence for the non-Gaussian halo bias with a slope similar to that of the local model. In the resonant case, we find that the non-Gaussian bias oscillates with halo mass, a novel feature that is unique to this type of models. In the case of a sharp feature in the inflaton potential, we find that the clustering of halos is enhanced at the mass scale corresponding to the Fourier mode that exited the horizon when the inflaton was crossing the feature in the potential. Both of these are new effects that open the possibility of characterizing the inflationary potential with large-scale-structure surveys. We briefly discuss the prospects for detecting these non-Gaussian effects. 

\end{abstract}
\maketitle

\section{Introduction}
Since the publication of the seminal inflation papers \cite{Guth:1980zm,Starobinsky:1980te}, a plethora of models have been proposed to explain why the Universe underwent a phase of exponential expansion at early times. Since most models offer very similar basic predictions, distinguishing between these models with today's data is not an easy task. One approach that has received a lot of attention recently is to look for departures from Gaussianity in the primordial cosmological perturbations \cite{Bernardeau:2001qr}. Indeed, while a large class of models predicts that the non-Gaussian signature should be undetectably small, there also exist a number of models for which departures from Gaussianity should be relatively large and observable \cite{Bartolo:2004if}.  Thus, any detection (or absence thereof) of non-Gaussianity in the primordial spectrum of perturbations could then rule out a large swath of inflation models.  

Non-Gaussian signatures have been traditionally looked for in cosmic microwave background (CMB) anisotropies \cite{Komatsu:2010fb}. However, it has recently been shown that the initial departure from Gaussianity could be amplified in the clustering of dark-matter halos \cite{Dalal:2007cu,Matarrese:2008nc,2008PhRvD..78l3507A} (see \cite{2010arXiv1006.4763D} for a review and \cite{Rubin:2001yw} for similar effects in another context). Indeed, mode coupling in non-Gaussian models induces a dependence of the local power spectrum on the long-wavelength potential perturbations. This can lead to a scale-dependent halo bias on large scales which is observable in large-scale-structure surveys, since galaxy clustering is closely connected to halo clustering on large scales.  Competitive upper limits on non-Gaussianity have already been placed using this method \cite{Slosar:2008hx}. 

While the non-Gaussian bias correction goes as $k^{-2}$ in the local model, it has been shown in \cite{2009ApJ...706L..91V,Schmidt:2010gw} that the scale dependence of other types of non-Gaussian models can be significantly different. Furthermore, models could also differ by how the bias varies with halo mass. Therefore, measurements of the biasing of dark-matter halos could be used to distinguish among different non-Gaussian scenarios. So far, the bispectrum shapes for which large-scale-structure predictions have been worked out include the local \cite{SalopekBond}, equilateral \cite{2006JCAP...05..004C}, orthogonal \cite{2010JCAP...01..028S}, and folded \cite{Meerburg:2009ys} shapes, all of which are scale independent.  However, there are several classes of inflationary models which predict bispectra that have strongly scale-dependent  oscillatory features \cite{Wang:1999vf,Chen:2006xjb,Chen:2008wn,Chen:2010xka,Chen:2011zf,Bean:2008na,McAllister:2008hb,Meerburg:2009ys,Flauger:2009ab,Meerburg:2009fi,Flauger:2010ja}.  
These models can circumvent the tight limit
on the bispectrum in the squeezed configuration \cite{Maldacena03} by 
breaking the slow-roll approximation.  Since the squeezed
triangle configuration is what determines the scale-dependent halo bias, 
such models potentially leave interesting signatures in halo clustering.  
The oscillatory bispectrum shapes are generally nonfactorizable and are therefore very computationally intensive to constrain with CMB data alone \cite{FLS,Meerburg:2010ks}. We show here that these models can also be constrained by calculating their impact on halo clustering. Moreover, these models leave distinct features in the mass dependence of the non-Gaussian halo bias, which allow us to distinguish them observationally from the smooth, scale-invariant shapes considered thus far. 

In this paper, we calculate the non-Gaussian correction to the dark-matter halo bias for two different oscillatory bispectra. We focus on models that display a sharp feature in their inflationary potential as well as models that have periodic features in the potential.  Models with features in the
potential have been invoked to explain deviations in the observed
CMB power spectrum from the smooth prediction.  On the other hand, periodic
modulations of the potential are motivated by axion-monodromy models \cite{McAllister:2008hb,Flauger:2009ab}.    
While in both cases one obtains oscillatory three-point functions, the physics responsible for these modulations is very different.  Indeed, the non-Gaussianities in the model with a feature are generated when the mode exits the horizon while for the resonant model, the non-Gaussianities are generated deep inside the horizon. As a consequence, we expect the two inflationary scenarios to make distinct  predictions about the clustering of dark-matter halos. In particular, we anticipate that in the feature model, the non-Gaussian effect should be the largest around the mass scale that exited the horizon while the inflaton was crossing the feature. On the other hand, we expect non-Gaussian effects to be important for a broad range of scales in the resonant model since these were generated by causal physics inside the horizon. Our results support these qualitative predictions and most interestingly, they allow us to map properties of the inflaton potential to features of galaxy clustering.

The structure of this paper is as follows. We begin by briefly reviewing halo biasing in the peak-background split formalism for general nonlocal quadratic non-Gaussianity. We then calculate the scale-dependent correction to the halo bias for the two inflation models considered here, emphasizing the effect of the new term unveiled in \cite{Desjacques:2011mq,Desjacques:2011jb}. We finally discuss our results in light of our qualitative predictions and physical expectations and conclude with a discussion on how these new effects could be detected in large-scale-structure data.

\section{Non-Gaussian Halo Bias in the Peak-Background Split Formalism}
\subsection{Nonlocal Kernel}
Following \cite{Schmidt:2010gw},  we consider the case for which the 
Bardeen potential during matter domination $\Phi$ is a general, 
nonlocal quadratic function of a Gaussian field $\phi$
\be\label{prim_pot_def}
\Phi(\xx)=\phi(\xx)+f_{\text{NL}}\int d^3y \int d^3z W(\yy,\zz)\phi(\xx+\yy)\phi(\xx+\zz),
\ee
where the kernel $W(\yy,\zz)$ is symmetric in its arguments and only depends on $y$, $z$ as well as $\hat{\yy}\cdot\hat{\zz}$. In Fourier space, one can think of $\widetilde{W}(\kk_1,\kk_2)$ as a scale-dependent coupling between different modes. To conform to standard notation, we have pulled out an arbitrary factor of $\fnl$ from the non-Gaussian kernel. Our results do not depend on this particular choice as they are only sensitive to the product $\fnl W(\yy,\zz)$. Deviation from Gaussianity is usually parametrized by the bispectrum,
\be
\langle\Phi(\kk_1)\Phi(\kk_2)\Phi(\kk_3)\rangle=(2\pi)^3\de^3_D(\kk_1+\kk_2+\kk_3)B_{\Phi}(k_1,k_2,k_3),
\ee
where $\de_D$ is the Dirac delta function and $k_3=|\kk_1+\kk_2|$. In terms of the Fourier space kernel $\widetilde{W}(\kk_1,\kk_2)$, the bispectrum amplitude is given by
\begin{align}\label{bispectrum}
B_{\Phi}(k_1,k_2,k_3)&=2f_{\text{NL}}\Big[\widetilde{W}(\kk_1,\kk_2)P_{\Phi}(k_1)P_{\Phi}(k_2)\en
& \qquad \qquad \qquad+  \text{2 perm.}\Big],
\end{align}
where $P_{\Phi}(k)$ stands for the power spectrum of $\Phi$. The two
permutations not written are the two remaining cyclic
permutations of $k_1$, $k_2$, $k_3$. Since the kernel $\widetilde{W}(\kk_1,\kk_2)$ is only required to be symmetric under the exchange of its two vectorial arguments, Eq. (\ref{bispectrum}) does not uniquely specify $\widetilde{W}$. One possible choice of kernel is 
\be
\widetilde{W}(\kk_1,\kk_2)=\frac{1}{2\fnl}\frac{B_{\Phi}(k_1,k_2,k_3)}{P_{\Phi}(k_1)P_{\Phi}(k_2)+2 \text{ perm.}},
\ee
which has the nice property of being fully symmetric under the exchange of the three momenta. For the halo bias calculation, we are mainly interested in the squeezed limit of the kernel where $k_2,k_3\gg k_1$. In this limit, the bispectrum uniquely defines the kernel via the relation \cite{Schmidt:2010gw}
\be\label{squeezed_W}
\widetilde{W}(\kk_1,\kk_2)\stackrel{k_2\gg k_1}{\longrightarrow}\frac{B_{\Phi}(k_1,k_2,k_3)}{4f_{\text{NL}}P_{\Phi}(k_1)P_{\Phi}(k_2)}.
\ee
Finally, to compute the dark-matter halo bias at late times, we need to consider the \emph{processed} kernel $\widetilde{W}_0(\kk_1,\kk_2)$ defined via the transfer function $T(k)$,
\be
\widetilde{W}_0(\kk_1,\kk_2)=\frac{T(|\kk_1+\kk_2|)\widetilde{W}(\kk_1,\kk_2)}{T(k_1)T(k_2)}.
\ee
In the squeezed limit, this reduces to
\be
\widetilde{W}_0(\kk_1,\kk_2)\stackrel{k_2\gg k_1}{\longrightarrow}\frac{1}{T(k_1)}\frac{B_{\Phi}(k_1,k_2,k_3)}{4f_{\text{NL}}P_{\Phi}(k_1)P_{\Phi}(k_2)}.
\ee
Note that we define our $\fnl$ in terms of the Bardeen potential at
last scattering, conforming to the convention usually adopted in CMB
analyses.  

\subsection{Halo Bias in Peak-Background Split}
We work in the Lagrangian picture of halo biasing where halos are
identified as high-density regions in the initial linear matter field. As such, we focus here on deriving the Lagrangian halo bias $b_I$ which relates the halo power spectrum to the linear matter power spectrum, $P_h(k)=b_I^2P(k)$. The late-time linear Eulerian bias relevant for observations on large scales is simply given by $b_1^{\rm E} = 1 + b_I$.
In the Lagrangian picture, the number density of halos per unit logarithmic mass (also called halo mass function) is sensitive to the statistics of small-scale perturbations. In the Gaussian case, each Fourier mode evolves independently and therefore the small-scale matter power spectrum $P(k_s)$ (at some initial early time) is the same everywhere. However, non-Gaussianity introduces mode coupling resulting in a dependence of the small-scale power spectrum on the local value of long-wavelength fluctuations. Non-Gaussian initial conditions thus generally rescale the local small-scale variance of the density field smoothed over a scale $R_s$, $\sigma_{0s}$, according to \cite{Desjacques:2011mq}
\be
\hat{\sigma}_{0s}^2\simeq\sigma_{0s}^2+4f_{\text{NL}}\phi_L(\kk)\sigma_W^2(k),
\label{eq:shat}
\ee
where the spectral moment $\sigma_W^2$ is
\be\label{sigma_w}
\sigma_W^2(k)=\int\frac{d^3k_s}{(2\pi)^3}F^2_{R_s}(k_s)\widetilde{W}_0(\kk,\kk_s)P(k_s),
\ee
and where $\phi_L(\kk)$ is a long-wavelength fluctuation of the gravitational potential. Here, $F_{R_s}$ is the Fourier transform of a spherical tophat with radius $R_s$, $P(k_s)$ is the matter power spectrum, and the ``hat'' denotes quantities that contain non-Gaussian contributions. Notice the appearance of the non-Gaussian kernel which indicates how a mode with wave number $k_s$ couples to the long-wavelength mode $k$.  Note also that $\sigma_W^2$ is not positive definite, as the
sign depends on the shape of the non-Gaussian kernel.  However, the
second term in Eq.~\ref{eq:shat} is always much smaller than the Gaussian
variance $\sigma_{0s}^2$ (since $\phi_L\sim 10^{-5}$), so that $\hat\sigma_{0s}^2$
is always positive.

Since the halo abundance $\hat{n}_h$ generically depends on $\hat{\sigma}_{0s}$, this induces a  scale-dependent dark-matter halo bias of the form \cite{Schmidt:2010gw,Desjacques:2011mq}
\begin{align}\label{bI}
b_I(M,z;k)&\equiv\left.\frac{1}{\hat{\bar{n}}_h}\frac{d\hat{\bar{n}}_h}{d\de_L(k)}\right|_{\de_L=0}\en
&=\frac{\pa\ln{\hat{\bar{n}}_h}}{\pa\ln{\rhob}}+\frac{\pa\ln{\hat{\bar{n}}_h}}{\pa\ln{\hat{\sigma}_{0s}}}\frac{\pa\ln{\hat{\sigma}_{0s}}}{\pa\de_L(k)}
\end{align}
where $M$ stands for the halo mass, $z$ for redshift, $\rhob$ is the average matter density of the Universe, $\hat{\bar{n}}_h$ is the average number density of halos of mass $M$ and $\de_L(k)$ is a long-wavelength density fluctuation. The halo mass $M$ is related to the smoothing scale $R_s$ through $M=(4\pi/3)\rhob R_s^3$ for a spherical tophat window function. In the following, we will drop the explicit $z$ dependence. The first term in Eq.~(\ref{bI}) is the usual Gaussian bias $b_1$ while the second term is induced by the non-Gaussian initial conditions. This last term can be expressed in a compact way when adopting a
universal mass function prescription,
\begin{equation}
\hat{\bar n}_h = \frac{\rhob}{M} f(\nu) \left|\frac{\partial\ln\hat{\s}_{0s}}{\partial\ln M}\right|,
\end{equation}
where $\nu = \delta_c/\hat{\sigma}_{0s}$ is the significance, 
$\de_c\approx1.686$ is the linearly extrapolated collapse threshold,
and $f(\nu)$ is a multiplicity function which we do not need to specify 
explicitly.  
A change in $\hat{\sigma}_{0s}$ thus changes halo abundance through a change
in $\nu$ as well as a change in the Jacobian $|\partial\ln\hat{\s}_{0s}/\partial\ln M|$.  The non-Gaussian halo bias correction can then be written in terms of the non-Gaussian kernel,
\begin{align}\label{bias1}
\Delta b_I(M,k)&=2f_{\text{NL}}\mathcal{M}^{-1}(k)\frac{\sigma_W^2(M,k)}{\sigma_{0s}^2(M)}\en
&\qquad\qquad\quad\times[b_1(M)\de_c+2\epsilon_{W}(M,k)],
\end{align}
with
\be
\epsilon_W(M,k)\equiv\frac{\pa\ln{\sigma_W^2(M,k)}}{\pa\ln{\sigma_{0s}^2(M)}}-1,
\ee
where $\mathcal{M}(k)=2k^2g_*(z)/(3(1+z)H_0^2\Omega_m)$. Here $g_*(z)$ is the potential growth function normalized to unity at last scattering. Since it is understood that $k_s\gg k$ in Eq.~(\ref{sigma_w}), we see that the bias correction depends on the non-Gaussian kernel evaluated in the squeezed limit. We note that the term proportional to $\epsilon_W(M,k)$ had been previously neglected in the literature until it was shown to be important in \cite{Desjacques:2011mq}. As we will see in the next section, this term is crucial for models displaying oscillatory features in their bispectrum. Examining Eq.~(\ref{bias1}), we observe that the scale dependence of the halo bias is determined by the product of $\mathcal{M}^{-1}(k)\propto k^{-2}$ with the leading $k$-dependent part of the processed non-Gaussian kernel evaluated in the squeezed limit. We now turn our attention to bispectra showing oscillatory behavior and calculate the resulting scale-dependent bias. The numerical results presented in this paper assume a flat $\Lambda$CDM universe with $h=0.72$, $\Omega_m=0.28$, $n_s=0.958$ and $\sigma_8=0.8$.  The pivot scale for the primordial power spectrum amplitude is kept at $k_* = 0.002 \rm Mpc^{-1}$ throughout.  

\section{Oscillatory Bispectra and their Scale-dependent Bias}
\subsection{Resonant Non-Gaussianity}
\begin{figure}[t]
\includegraphics[width=0.5\textwidth]{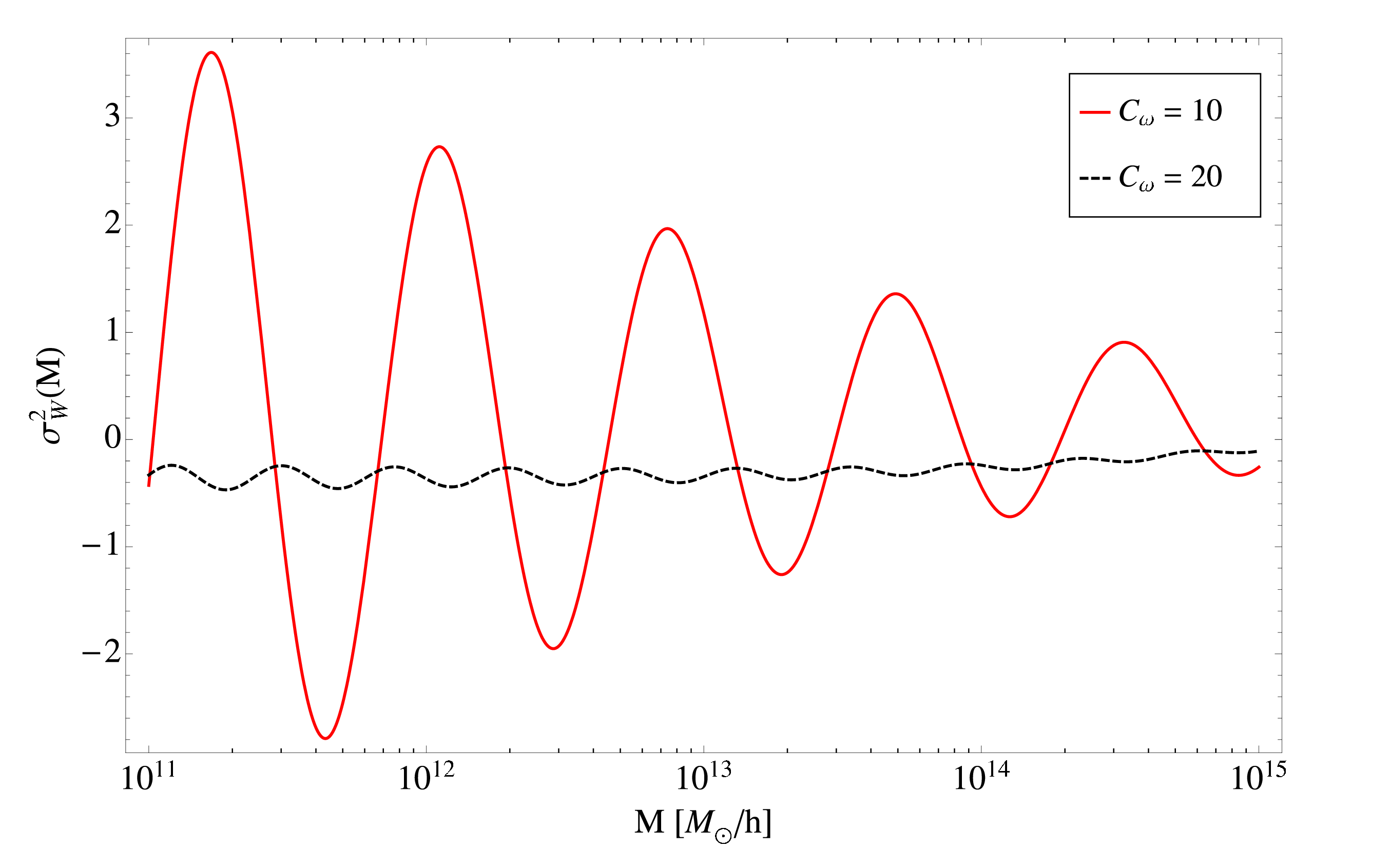}
\caption{Non-Gaussian spectral moment $\sigma_W^2$  for the resonant model as a function of halo mass. We evaluate this spectral moment for $k=10^{-3}h$ Mpc$^{-1}$ at $z=0$. }
\label{sigmaWvsM_res}
\end{figure}

Resonant non-Gaussianity arises when periodic features in the inflationary potential lead to an oscillatory coupling between modes, which can trigger a resonance for modes oscillating with the same frequency inside the horizon \cite{Chen:2008wn,Chen:2010xka,Bean:2008na,McAllister:2008hb,Flauger:2009ab,Flauger:2010ja}. Such features arise, for example, in certain brane inflation models or in axion-monodromy inflation.  For this class of models, the bispectrum has the generic form \cite{Flauger:2010ja}
\begin{align}\label{bi_res}
B_{\text{res}}&=\left(\frac{5}{3}\right)(2\pi)^4f^{\text{res}}_{\text{NL}}\Delta_{\Phi}^2\frac{1}{k_1^2k_2^2k_3^2}\Bigg(\sin{(C_{\omega}\ln{(k_t/k_p)})}\en
&+\frac{1}{C_{\omega}}\cos{(C_{\omega}\ln{(k_t/k_p)})}\sum_{i\neq j}\frac{k_i}{k_j}+\mathcal{O}\left(\frac{1}{C_\omega^2}\right)\Bigg),
\end{align}
where $\Delta_\Phi$ is the amplitude of primordial scalar power spectrum, $k_t=k_1+k_2+k_3$, $k_p$ is a pivot scale which introduces a phase, and $C_{\omega}$ is related to the frequency $\omega$ of the periodic features of the inflationary potential by $C_{\omega}=\omega/H_I$. Here, $H_I$ stands for the Hubble parameter during inflation. The leading factor of $5/3$ comes from the conversion between the Bardeen primordial potential $\Phi$ and the gauge invariant curvature perturbation $\zeta$ at late times. Constraints from the matter power spectrum provide an upper bound on the value of $f^{\text{res}}_{\text{NL}}$. For the axion-monodromy scenario with a linear zero-order potential, this bound reads \cite{Flauger:2010ja,Meerburg:2010ks}
\be\label{upper_bound_fnl}
f^{\text{res}}_{\text{NL}}\lesssim10^{-3}C_{\omega}^{5/2},
\ee
where it is assumed that the pivot scale $k_p=0.002$ Mpc$^{-1}$ exits the horizon about 60 e-folds before the end of inflation. Other zeroth-order inflaton potentials are likely to lead to a somewhat different constraint on $f^{\text{res}}_{\text{NL}}$ but we shall use Eq.~(\ref{upper_bound_fnl}) as a rough upper limit for this type of model. The resonant bispectrum can readily be evaluated in the squeezed limit \cite{Flauger:2010ja}
\be
B_{\text{res}}\stackrel{k_s\gg k}{\longrightarrow}\left(\frac{5}{3}\right)\frac{(2\pi)^4f^{\text{res}}_{\text{NL}}}{C_{\omega}}\frac{2\Delta_\Phi^2}{k_s^3k^3}\cos{\left\{C_{\omega}\ln{\left[\frac{2k_s}{k_p}\right]}\right\}}.
\ee
Note that the leading correction to this expression is suppressed by a factor of $k/k_s$, thus negligibly contributing to the bispectrum in the squeezed limit. Using Eq.~(\ref{squeezed_W}), we obtain the leading-order non-Gaussian kernel, 
\begin{align}\label{kernel_res_sq}
\widetilde{W}(\kk,\kk_s)&\simeq\left(\frac{5}{3}\right)\frac{(2\pi)^4}{2C_{\omega}}\left(\frac{k}{k_*}\right)^{-\epsilon}\cos{\left\{C_{\omega}\ln{\left[\frac{2k_s}{k_p}\right]}\right\}}\en
&\qquad\times\left(\frac{k_s}{k_*}\right)^{-\epsilon}\left[1-\frac{1}{2}\left(\frac{k}{k_s}\right)^{3-\epsilon}+\ldots\right],
\end{align}
where $\epsilon=n_s-1$ and the ellipsis stands for terms that are higher order in $k/k_s$.
We immediately see that the scale dependence of the non-Gaussian bias is given by
\be
\Delta b_{I,\rm res}(k)\propto k^{-2-\epsilon},
\ee
that is, it is very similar to that of the local model. To calculate the amplitude of the bias correction, we first need to integrate Eq.~(\ref{sigma_w}) over the small-scale modes to obtain the non-Gaussian spectral moment $\sigma_W^2(k)$. The $k_s$ integral is of the general form
\be
\sigma^2_W\propto\int k_s dk_s T^2(k_s)\frac{j_1^2(k_sR_s)}{R_s^2}\cos{(C_{\omega}\ln{(2k_s/k_p)})}.
\ee
For large values of the frequency $C_{\omega}$, the integrand is rapidly oscillating and the resulting amplitude for the non-Gaussian bias is expected to be rather small. For a small enough value of the frequency ($C_{\omega}\lesssim100$), the integral can be done numerically. To evaluate the second term in Eq.~(\ref{bias1}), we first use the chain rule to write it as
\be
\frac{\pa\sigma_W^2(M,k)}{\pa\sigma_{0s}^2(M)}=\frac{\pa\sigma_W^2(M,k)}{\pa\ln M}\left(\frac{\pa\sigma_{0s}^2(M)}{\pa\ln M}\right)^{-1}.
\ee
\begin{figure}[t]
\includegraphics[width=0.5\textwidth]{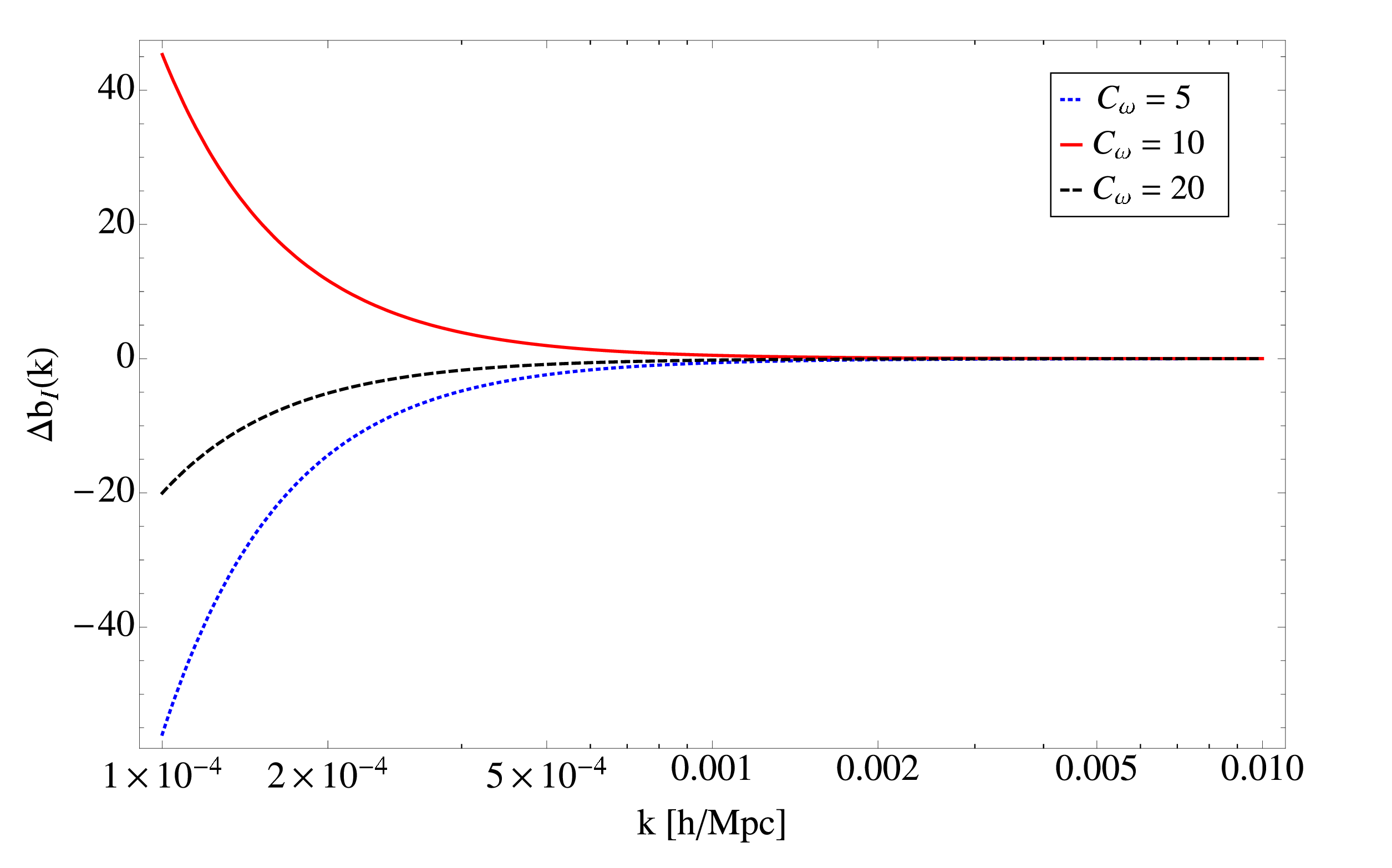}
\caption{Non-Gaussian correction to the halo bias for the resonant non-Gaussianity model as a function of scale. We evaluate the bias for $M=10^{13}M_{\odot}/h$ at $z=0$. We take $f_{\text{NL}}^{\text{res}}=10^{-3}C_{\omega}^{5/2}$ and evaluate the Gaussian bias $b_1$ using the Sheth-Tormen mass function \cite{ShethTormen}.}
\label{bias_vs_k_res}
\end{figure}
The derivatives on the right can be calculated numerically. For all numerical computations, we use the complete expression for the bispectrum, Eq.~(\ref{bi_res}). In Fig.~\ref{bias_vs_k_res}, we show the scale dependence  of the non-Gaussian halo bias correction for three values of $C_{\omega}$ evaluated for a halo mass of $10^{13}M_{\odot}/h$ at $z=0$. We see that the non-Gaussian bias correction is small except for the largest scales where the scale dependence of $\Delta b_{I,\rm res}(k)\propto k^{-2-\epsilon}$ becomes important. Interestingly, the non-Gaussian bias for these resonant models is completely dominated by the term proportional to $\epsilon_W(M,k)$ in Eq.~(\ref{bias1}), which was recently unveiled in \cite{Desjacques:2011mq,Desjacques:2011jb}. To understand why this new term is crucial for our analysis, we plot in Fig.~\ref{sigmaWvsM_res} the non-Gaussian spectral moment $\sigma_W^2$ as a function of halo mass for a fixed comoving scale. We see that $\sigma_W^2$ strongly oscillates with halo mass, leading to a large contribution to $\pa\sigma_W^2/\pa\ln M$, especially toward small masses. This highlights the importance of the newly discovered term for accurately predicting the non-Gaussian halo bias.  We will discuss the relevance of this result for observations in Sec.~\ref{sec:disc}.  

\begin{figure}
\includegraphics[width=0.5\textwidth]{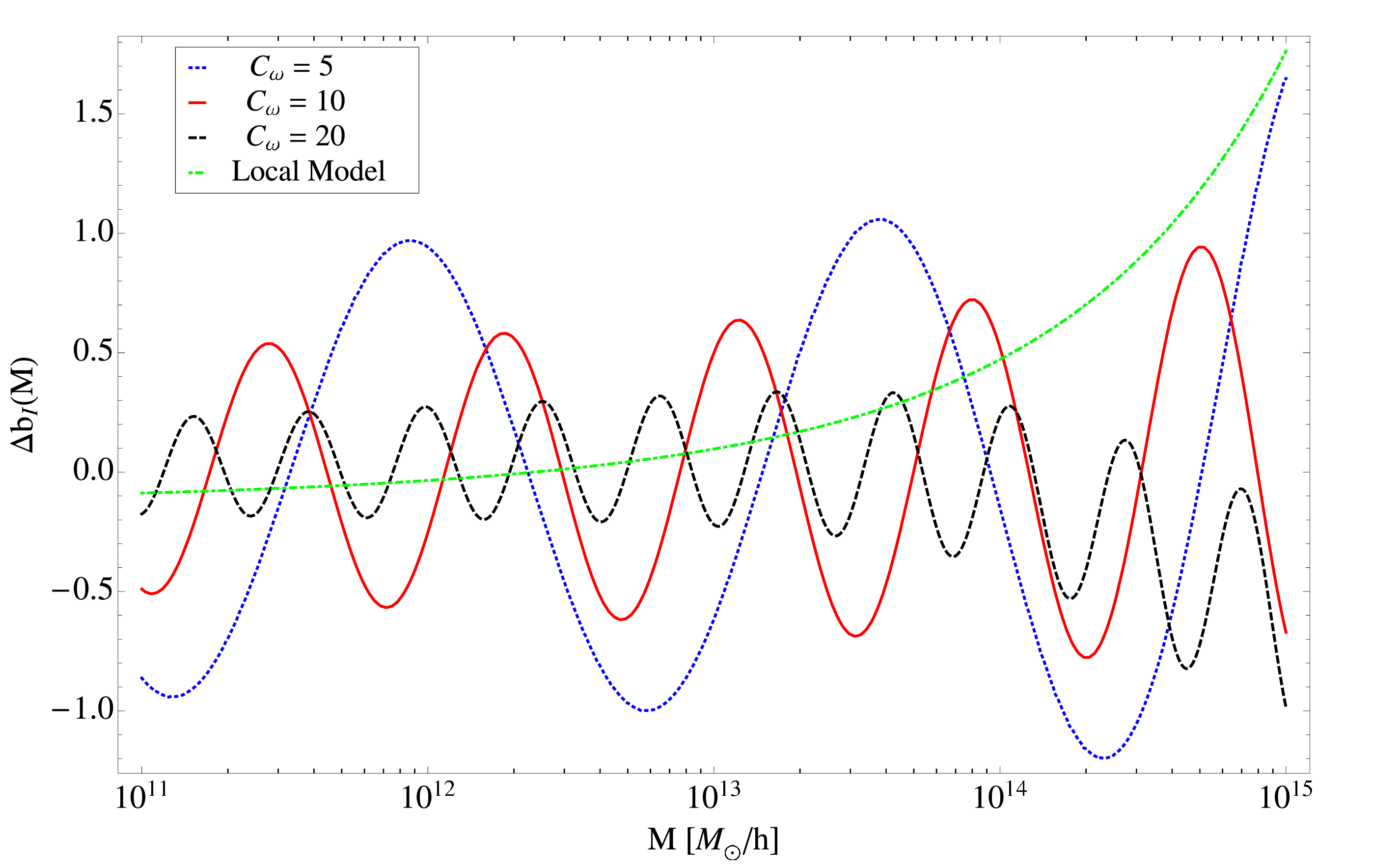}
\caption{Non-Gaussian correction to the halo bias for the resonant non-Gaussianity model as a function of halo mass. We evaluate the bias for $k=10^{-3}h$ Mpc$^{-1}$ at $z=0$. We take $f_{\text{NL}}^{\text{res}}=10^{-3}C_{\omega}^{5/2}$ and evaluate the Gaussian bias $b_1$ using the Sheth-Tormen mass function. For comparison, we also show the bias for local non-Gaussianity with $f_{\text{NL}}^{\text{local}}=2$.}
\label{bias_vs_M_res}
\end{figure}

An interesting feature of resonant non-Gaussianity models is that they predict a modulation of the halo bias with changing halo mass. In Fig.~\ref{bias_vs_M_res}, we show  the non-Gaussian halo bias as a function of halo mass evaluated at a scale $k=10^{-3}h$ Mpc$^{-1}$. Again, the bias is dominated by the second term of Eq. (\ref{bias1}) for $M\lesssim10^{15}M_{\odot}/h$. We observe that the amplitude of the non-Gaussian bias decreases with increasing $C_{\omega}$ very rapidly and therefore this effect is likely to be unobservable unless $C_{\omega}$ is small. As expected, the non-Gaussian features of the halo bias show coherent modulations over a wide range of mass scales, an artifact of non-Gaussianities being produced by causal physics deep inside the horizon for these models. To contrast the resonant model with the more traditional local model of non-Gaussianity, we also display the halo bias for a local model with $f_{\text{NL}}^{\text{local}}=2$.  At moderate halo masses ($M\!\sim\!10^{12}\!-\!10^{14} M_\odot/h$), the effect of resonant non-Gaussianity is comparable to that of a local model with $\fnl^{\rm local}$ of order unity, for the values of $C_\omega$ chosen.  

\subsection{Features in the Inflaton Potential}
The presence of a sharp feature in the inflaton potential can induce large primordial non-Gaussianities \cite{Wang:1999vf,Chen:2006xjb,Chen:2008wn,Chen:2010xka,Chen:2011zf}. Indeed, modes that exit the horizon while the inflaton is crossing the feature get a boost in their three-point signal. Here, we shall focus on the case of a step in the inflaton potential, but our analysis could also be applied to the case of a bump in the potential.  The exact form of the bispectrum can only be obtained numerically but the authors of \cite{Chen:2008wn} suggested an approximate form:
\begin{align}\label{bi_feat_full}
B_{\text{feat}}(k_1,k_2,k_3)&\approx-\left(\frac{5}{3}\right)(2\pi)^4\fnl^{\text{feat}}\frac{\Delta_\Phi^2k_f}{k_1^3k_2^3k_3^3}\\
&\times\left\{2\frac{\sum_{i\neq j}k_ik_j^2}{k_t}\sin{\left[\frac{k_t}{k_f}\right]}\sin{\left[\frac{k_t\Delta k_f}{k_f^2}\right]}\right\}\nonumber.
\end{align}
Here, $k_t=k_1+k_2+k_3$ and $k_f\pm\Delta k_f$ are the Fourier modes that exit the horizon while the inflaton is crossing the feature (the sharper the feature in the inflaton potential, the larger $\Delta k_f$ becomes).  We fixed the phases such that $B_{\text{feat}}\rightarrow0$ as $k_t\rightarrow0$ which is physically motivated since modes that exit the horizon long before the inflaton encounters the feature should not show significant non-Gaussianities.  We choose the overall sign such that the non-Gaussian bias is positive for the scale exiting the horizon when the inflaton crosses the feature. In practice, this sign should be fixed by comparison to numerical simulations. In the squeezed limit, the bispectrum reads
\begin{align}\label{bi_feat}
B_{\text{feat}}&\stackrel{k_s\gg k}{\longrightarrow}-\left(\frac{5}{3}\right)(2\pi)^4\fnl^{\text{feat}}\frac{\Delta_\Phi^2}{k_s^4k^2}\left(\frac{k_f}{k}\right)\en
&\qquad\qquad\times\sin{\left(\frac{2k_s}{k_f}\right)}\sin{\left(\frac{2k_s\Delta k_f}{k_f^2}\right)}.
\end{align}
For a narrow feature, we generally expect $\Delta k_f/k_f\ll1$ and thus the last sinusoidal factor in Eq.~(\ref{bi_feat}) can be considered as an envelope function for the first rapidly oscillating sine factor. From Eq.~(\ref{bi_feat}), we see that at fixed $k_s$, non-Gaussianity becomes more important for modes $k$ smaller than $k_f$ (remember that $k$ is the scale at which clustering of halos is measured). The non-Gaussian kernel can then readily be obtained as
\begin{align}\label{kernel_feat}
\widetilde{W}(\kk,\kk_s)&\simeq-\left(\frac{5}{3}\right)\frac{(2\pi)^4}{4}\left(\frac{k_f}{k_s}\right)\left(\frac{k}{k_*}\right)^{-\epsilon}\left(\frac{k_s}{k_*}\right)^{-\epsilon}\en
&\qquad\qquad\times\sin{\left(\frac{2k_s}{k_f}\right)}\sin{\left(\frac{2k_s\Delta k_f}{k_f^2}\right)}.
\end{align}
Note that $\widetilde{W}$ approaches zero for $k_s \ll k_f$ and for $k_s\gg k_f$.  This physically makes sense since modes with $k_s\gg k_f$ are oscillating deep inside the horizon when the inflaton crosses the feature and we expect their non-Gaussianities to roughly cancel out. On the other hand, modes with $k_s\ll k_f$ are outside the horizon when non-Gaussianities are generated and we thus expect their contribution to the kernel to be small. We immediately see that the scale dependence of the halo bias is given by
\be
\Delta b_{I,\rm feat}(k)\propto k^{-2-\epsilon},
\ee
which, at first look, is similar to the resonant model. However, as we will see below, the two models predict very different behaviors for how the amplitude of $\Delta b_I$ varies with halo mass. From Eq.~(\ref{kernel_feat}), it is straightforward to compute numerically the non-Gaussian spectral moment $\sigma_W^2$. In the feature model, the integral over small scale modes has the general form
\begin{align}
\sigma^2_W&\propto \int k_s dk_sT^2(k_s) \frac{j_1^2(k_sR_s)}{R_s^2} \left(\frac{k_f}{k_s}\right)\en&\qquad\qquad\qquad\times\sin{\left(\frac{2k_s}{k_f}\right)}\sin{\left(\frac{2k_s\Delta k_f}{k_f^2}\right)}.
\end{align}
We see that for a given value of $k_f$, there will always be a scale $R_s$ for which constructive interference between the first sine factor and the Bessel function happens (again, the second sine factor is considered to be slowly varying). Since the Bessel function peaks around $k_s\sim R_s^{-1}$, we naively expect $|\sigma_W^2|$ to have a maximum near $R_s=R_f\sim\mathcal{O}(1)/k_f$ or equivalently, near $M_f\sim\mathcal{O}(1)(4\pi/3)\rhob k_f^{-3}$. For $R_s\ll R_f$, the first sine factor is rapidly oscillating near the peak of the Bessel function and thus the only nonvanishing contribution comes from the low $k_s$ tail. In this limit, the integrand becomes independent of $R_s$ since $j_1^2(k_sR_s)\propto k^2_sR^2_s$ for $k_s\ll R_s^{-1}$. We thus expect $\sigma_W^2$ to asymptote to a constant for small halo masses.  On the other hand, for $R_s\gg R_f$, the integrand approaches zero and we therefore expect $\sigma_W^2$ to vanish for large halo masses.
\begin{figure}
\includegraphics[width=0.5\textwidth]{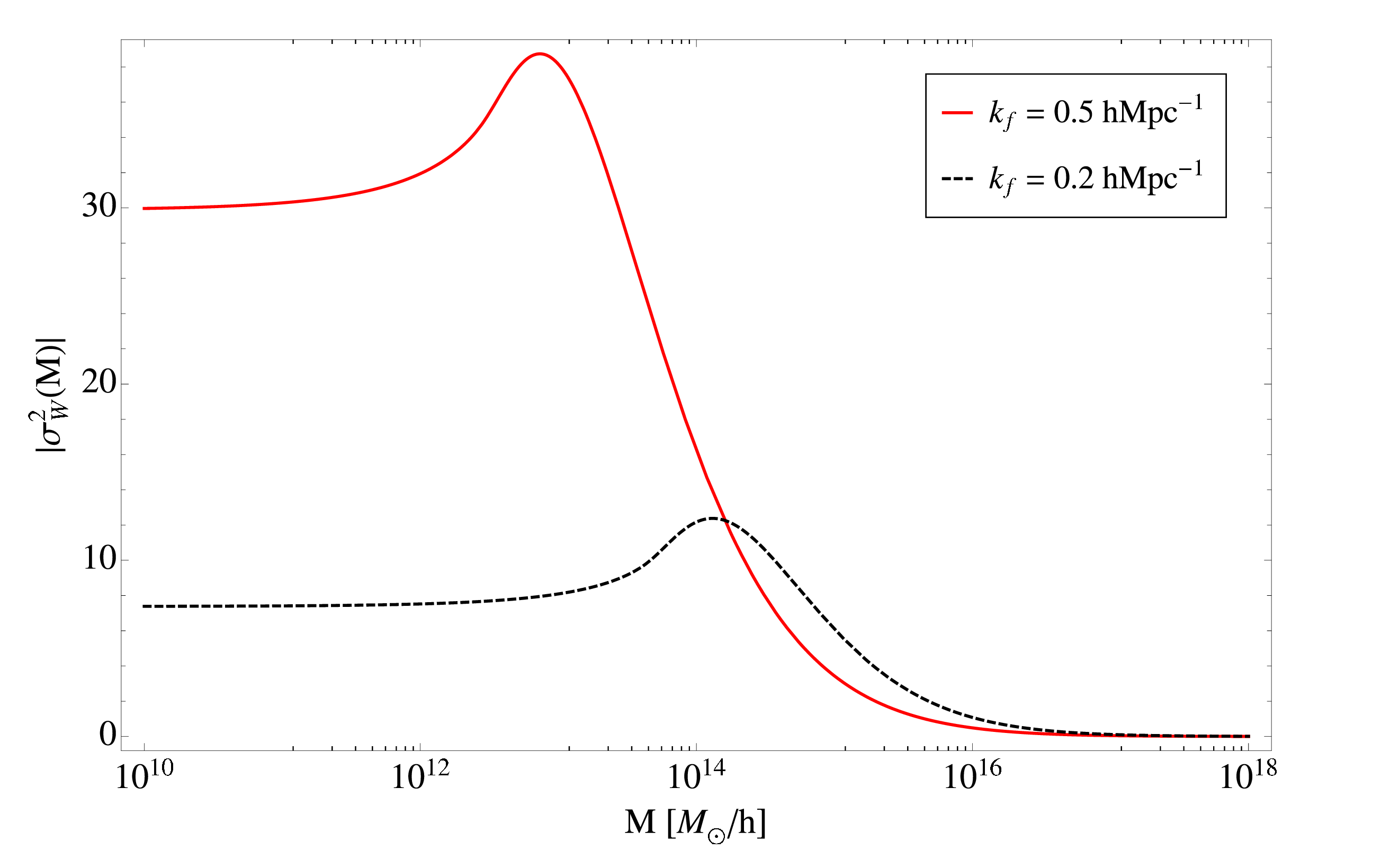}
\caption{Absolute value of the non-Gaussian spectral moment $\sigma_W^2$  for the feature model as a function of halo mass. We evaluate this spectral moment for $k=10^{-3}h$ Mpc$^{-1}$ at $z=0$ and use $\Delta k_f/k_f=0.01$. }
\label{feat_sigmaW_vsM}
\end{figure}

Our numerical calculations support these qualitative conclusions. Indeed, Fig.~\ref{feat_sigmaW_vsM} shows the absolute value of the non-Gaussian spectral moment for two values of $k_f$. The enhancement around $M_f\sim1.5(4\pi/3)\rhob k_f^{-3}$ is clearly visible in both cases and $\sigma_W^2(M)$ rapidly vanishes for $M>M_f$. The latter point is expected on physical grounds since halos with $M>M_f$ correspond to scales (in the initial density field) that were outside the horizon when the non-Gaussianities were generated. Therefore, the non-Gaussian correction to the variance of the density field smoothed on these scales must be somewhat suppressed.

\begin{figure}
\includegraphics[width=0.5\textwidth]{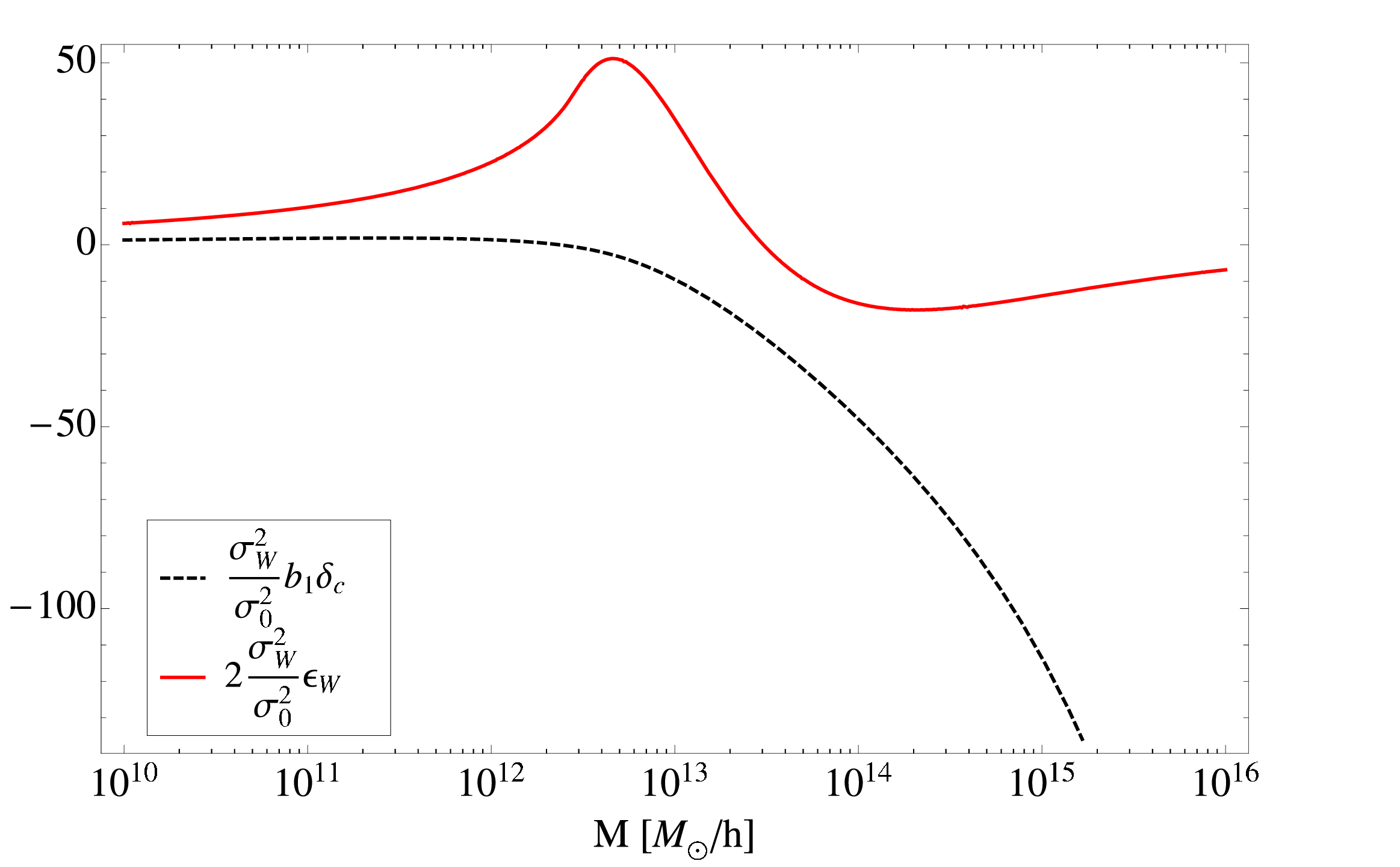}
\caption{The two contributions to the non-Gaussian halo bias correction as a function of mass for the model with a feature at $k_f=0.5$hMpc$^{-1}$. We evaluate $\sigma_W^2$ for $k=10^{-3} $hMpc$^{-1}$ at $z=0$ and use $\Delta k_f/k_f=0.01$. The Gaussian bias $b_1$ is derived using the Sheth-Tormen mass function.}
\label{feat_2terms}
\end{figure}

We can now use Eq.~(\ref{bias1}) to compute the scale-dependent non-Gaussian correction to the halo bias. From the functional form of $\sigma_W^2$, we expect the term proportional to $\epsilon_W$ to dominate around the feature at $M=M_f$ since $\pa\ln{\sigma_W^2}/\pa\ln{M}$ is largest there. To verify this, it is instructive to consider the two distinct contributions to the bias as a function of halo mass. In Fig. \ref{feat_2terms}, we show both the contribution proportional to the Gaussian bias $b_1$ as well as the recently unveiled contribution proportional to $\epsilon_W$ for a feature at $k_f=0.5$ hMpc$^{-1}$. We observe that the $\epsilon_W$ term clearly displays a feature at $M_f\sim1.5(4\pi/3)\rhob k_f^{-3}\simeq4\times10^{12}M_{\odot}/h$ and that it dominates the overall bias for halo masses $M\lesssim10^{14}M_{\odot}/h$. This once again highlights the importance of the $\epsilon_W$ term for models with strongly scale-dependent bispectra \cite{Desjacques:2011jb}.

\begin{figure}
\includegraphics[width=0.5\textwidth]{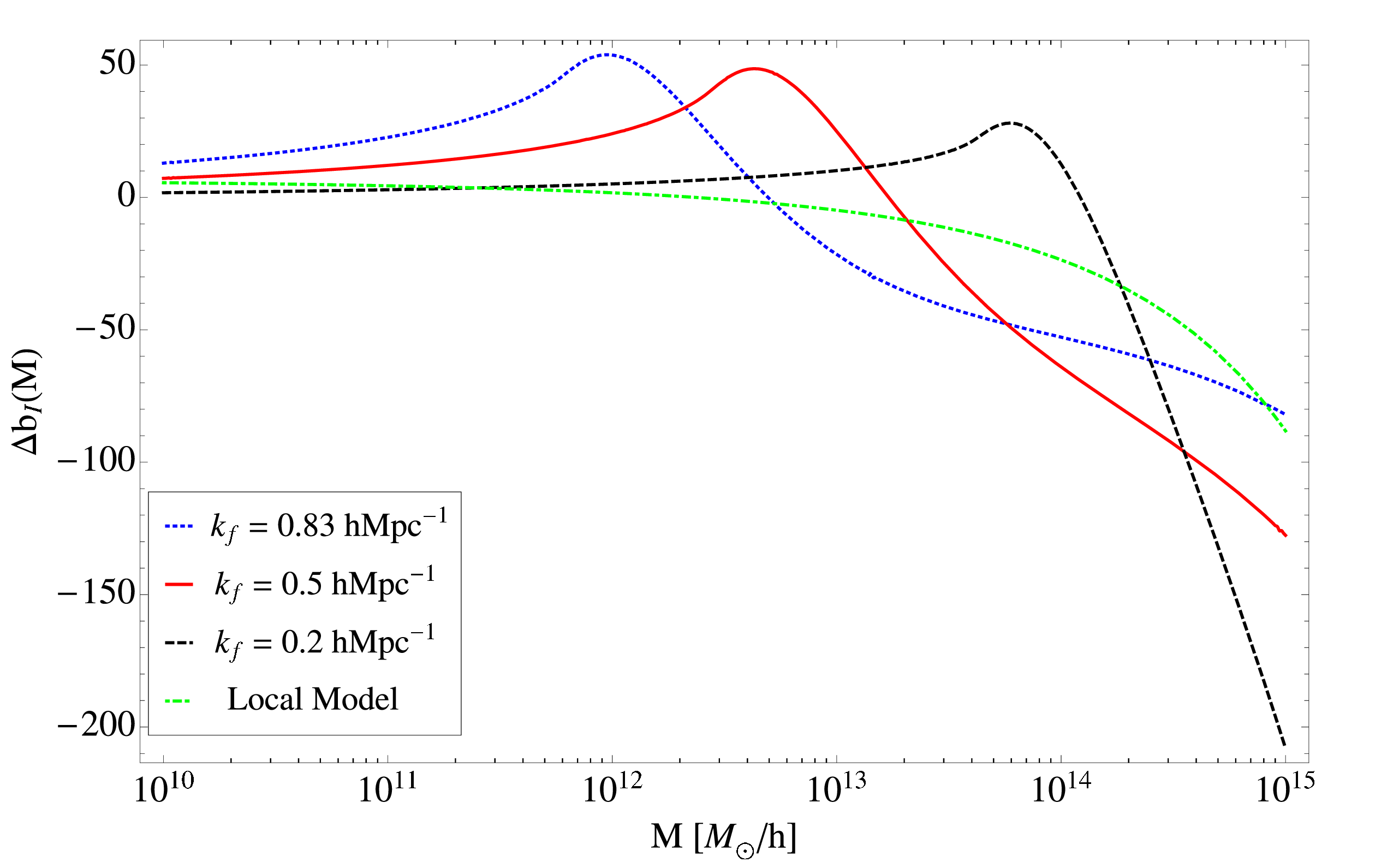}
\caption{Mass dependence of the non-Gaussian correction to the halo bias for the model with a feature in the inflaton potential. We evaluate the bias for $k=10^{-3}h$ Mpc$^{-1}$ at $z=0$ and use $\Delta k_f/k_f=0.01$. We take $f_{\text{NL}}^{\text{feat}}=10$ and evaluate the Gaussian bias $b_1$ using the Sheth-Tormen mass function. For comparison, we also show the bias for the local model of non-Gaussianity with $f_{\text{NL}}^{\text{local}}$=-10.}\label{bias_vs_M_feat}
\end{figure}

In Fig.~\ref{bias_vs_M_feat}, we show the amplitude of the non-Gaussian bias correction as a function of halo mass for three different values of $k_f$.  Note that while we assumed a value of $f_{\rm NL}^{\rm feat} = 10$ here, still within the allowed range of power spectrum contraints \cite{Chen:2008wn}, this parameter is in reality fixed for a given inflation potential by the numerical calculation.  For comparison, we also plot the halo bias correction for local quadratic non-Gaussianity, $\Delta b_{I,\rm local}=2f_{\text{NL}}^{\text{local}}b_1\de_c\mathcal{M}^{-1}(k)$. We immediately see that the bias correction for models with a feature in the potential displays a large enhancement around $M=M_f$ when compared to the monotonic and featureless bias of a local-type model. Observations of this tell-tale signature in large-scale-structure data provide us with an exciting new window to probe microscopic inflationary physics with observations of galaxy clustering on the largest scales. 

For halo masses $M\gg M_f$, the bias becomes dominated by the first term of Eq.~(\ref{bias1}) since $b_1$ is large for very massive halos. In this limit, $\Delta b_I$ is very sensitive to the small $k$ behavior of the bispectrum which may not be accurately captured by our ansatz Eq.~(\ref{bi_feat}). Thus, a complete numerical computation of the bispectrum is likely to be required to accurately predict the large-mass limit of the halo bias. We leave this for future work. However, since halos above $M\gtrsim 10^{15} M_\odot/h$ are very rare, especially at higher redshifts, we do not expect the observational constraints to be dominated by this mass range. 

\begin{figure}
\includegraphics[width=0.5\textwidth]{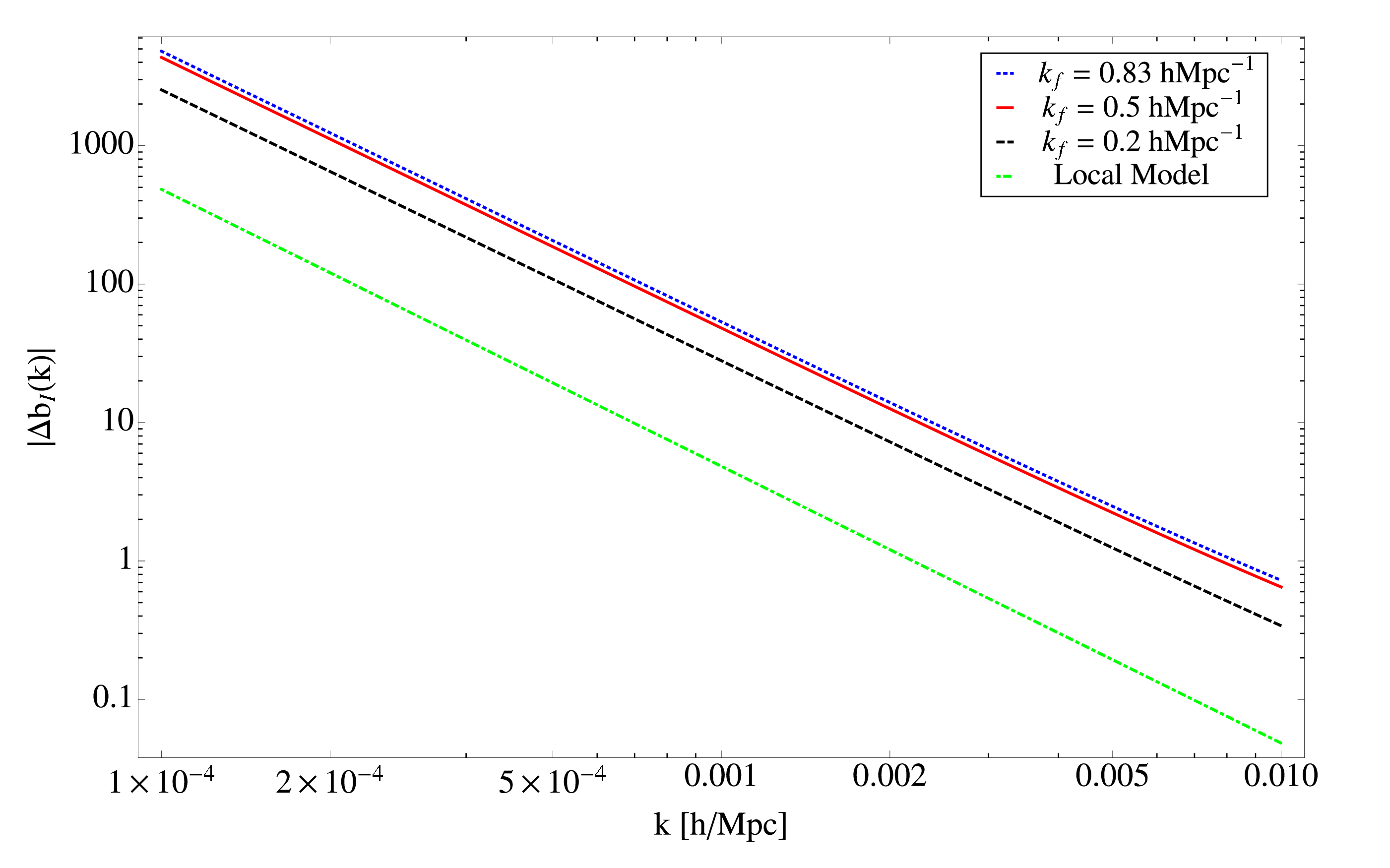}
\caption{Scale dependence of the non-Gaussian correction to the halo bias for the model with a feature in the inflaton potential. We evaluate the bias for $M=1.5(4\pi/3)\rhob k_f^{-3}$ at $z=0$ and use $\Delta k_f/k_f=0.01$. We take $f_{\text{NL}}^{\text{feat}}=10$ and evaluate the Gaussian bias $b_1$ using the Sheth-Tormen mass function. For comparison, we also show the non-Gaussian bias correction from the local model evaluated at $M=10^{13}M_{\odot}/h$ and $\fnl^{\text{local}}=100$.}
\label{bias_vs_k_feat}
\end{figure}
In Fig. \ref{bias_vs_k_feat}, we show the absolute value of the non-Gaussian bias correction as a function of scale for three values of $k_f$. We evaluate the amplitude of the bias at $M=M_f$, that is, at the peak of the feature in $\epsilon_W$. At this mass scale, we see that the halo bias can reach a very wide amplitude on large cosmological scales. The scale dependence proportional to $k^{-2-\epsilon}$ derived in Eq.~(\ref{kernel_feat}) is readily visible. For comparison, we also plot the halo bias correction for the local-model of non-Gaussianity which displays a similar scale dependence but a much smaller amplitude even for  $\fnl^{\text{local}}=100$. 

We note in passing that varying $\Delta k_f/k_f$ corresponds to changing the overall scale of the bispectrum. This can readily be seen from Eq.~(\ref{bi_feat}) where $\Delta k_f/k_f$ only appears in the modulating envelope. For $\Delta k_f/k_f\ll1$ and $2k_s\lesssim k_f$ (i.e. for the modes that contribute most to $\sigma_W^2$ ), one can Taylor expand the second sine factor to obtain, after simplification, $B_{\text{feat}}\propto\Delta k_f/k_f$. Therefore, the overall amplitude of the bias correction is determined by the product $f_{\text{NL}}^{\text{feat}}(\Delta k_f/k_f)$. This scaling agrees with the result of \cite{Chen:2006xjb}, where it was shown that the overall amplitude of the bispectrum is inversely proportional to the width of the step in the potential. To see this, we note that the sharper is the step, the more kinetic energy is acquired by the inflaton and by consequences, the longer slow roll is violated. Therefore, we expect that the sharper is the step (corresponding to larger bispectrum amplitude), the larger the band of Fourier modes affected ($\Delta k_f$) will be, hence the above result.

In summary, we have shown that the presence of a feature in the inflaton potential leads to a corresponding feature in the mass dependence of the non-Gaussian halo bias. Ultimately, this is a consequence of non-Gaussianity being generated at a specific scale during inflation in these models.  Finally, we reiterate that the numerical results presented in this section were computed using the analytical expression for the bispectrum given in Eq.~\ref{bi_feat_full}.  It is important to keep in mind that this expression is approximate.  However, it most likely captures the important physics. As such, we expect our conclusions to be robust to the inclusion of a more accurate bispectrum. 

\section{Discussion}
\label{sec:disc}
We have analyzed the non-Gaussian halo bias resulting from two inflation models displaying oscillatory bispectra. Even though the two models predict the same scale dependence as local quadratic non-Gaussianity, we find that they make very different predictions concerning how the amplitude of the bias varies with halo mass. Indeed, while the resonant model predicts an oscillatory amplitude as a function of halo mass, models with a feature in the potential predict an enhancement of the bias for halos with mass that corresponds to the scale that exited the horizon at the time when the inflaton was crossing the feature in the potential. Ultimately, these very different outcomes can be traced back to the distinct physics that is responsible for generating non-Gaussianities in the first place.\\

For the resonant model, non-Gaussianity is generated well inside the horizon when the modes are rapidly oscillating. As explained in \cite{Chen:2008wn}, oscillations in the inflaton potential lead to an oscillatory coupling between different Fourier modes. As the physical frequency of each mode $k/a(t)$ decreases, there will be a time when $k/a(t_{res})\sim\omega$ and the oscillating mode can resonate with the coupling and generate a departure from Gaussianity. As a large number of modes eventually passes through the resonance, we naturally expect the non-Gaussian effects to be present on a broad range of scales, as can be seen in Fig.~\ref{bias_vs_M_res}. A crucial consequence of this subhorizon generation mechanism is that the resulting modulation of the halo bias is \emph{in phase} over a broad range of masses. This is a tell-tale signature that could be looked for in large-scale-structure data and used to put constraints on these resonant-type inflation models.\\

For models with a feature in the inflaton potential, non-Gaussianity is generated during slow-roll violation associated with the inflaton suddenly accelerating as it crosses the step or the bump in the potential.  Consequently, different Fourier modes within a limited range of scales are coupled, hence generating a nonvanishing three-point function. However, modes deep inside the horizon are rapidly oscillating and we thus expect their non-Gaussian signature to average out to zero. On the other hand, modes that exit the horizon as slow roll is violated are frozen-in before causal physics could erase their correlation with other Fourier modes. We thus expect the bispectrum to be significant when at least one side of the triangle has $k\sim k_f$ (and no side with $k\gg k_f$). As a consequence, modes that exit the horizon during slow-roll violation get an enhanced coupling to the long-wavelength perturbations resulting in an amplified clustering of halos at the corresponding mass scale. Conversely, modes that are superhorizon when slow-roll is violated become correlated with modes that have $k\sim k_f$. This induces a rescaling of the variance of the density field according to Eq.~(\ref{eq:shat}) which results in a nonvanishing halo bias at these mass scales. As mentioned earlier, this rescaling of the variance is very sensitive to the small-k limit of the bispectrum and a complete numerical computation will be required to accurately predict the halo bias for $M\gg M_f$. Nevertheless, since very massive halos are rare and restricted to low redshifts, it is unlikely that observational constraints will depend sensitively on the high-mass tail.\\
\begin{figure}[t!]
\includegraphics[width=0.5\textwidth]{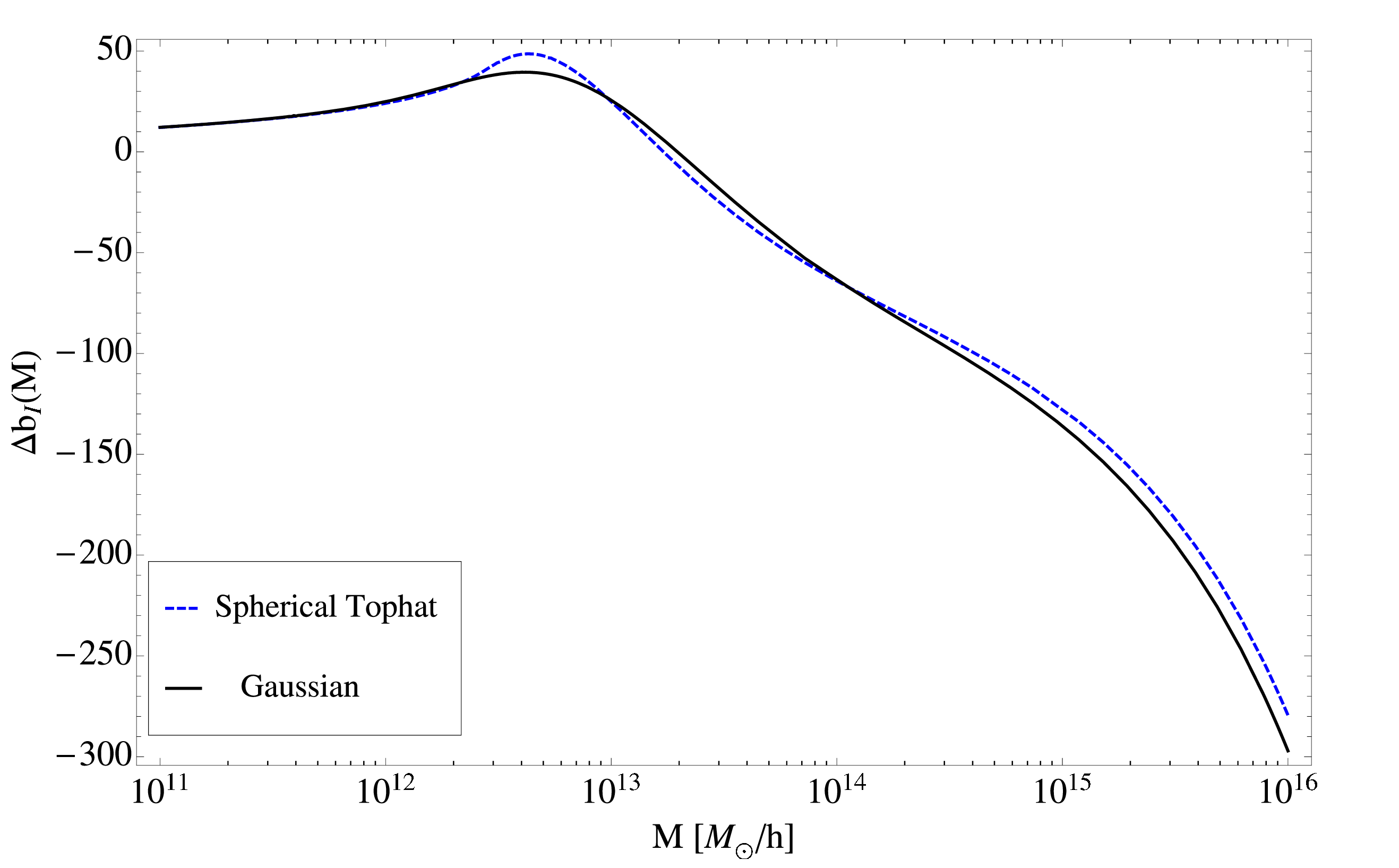}
\caption{Comparison between $\Delta b_I(M)$ obtained with a spherical tophat filter and with a Gaussian filter. We evaluate the bias for $k=10^{-3}h$ Mpc$^{-1}$ at $z=0$ and use $\Delta k_f/k_f=0.01$ and $k_f=0.5h$ Mpc$^{-1}$. We take $f_{\text{NL}}^{\text{feat}}=10$.}
\label{tophat_vs_gaussian}
\end{figure}

For both classes of models, we find that the term coming from the
Jacobian $d\ln\sigma_{0s}/d\ln M$ (see \cite{Desjacques:2011mq} for details) is an important and often dominant
contribution to the non-Gaussian halo bias.  This term comes about
because the non-Gaussian mode-coupling induces a modulation with $\phi_L$ of 
the significance \emph{interval} $d\ln\nu = d\ln\s_{0s}$ that corresponds to a 
fixed logarithmic mass interval $d\ln M$.  This term is strictly present because
we have assumed throughout that halos are selected by mass, which is 
appropriate e.g. when comparing to $N$-body simulations.  In practice however,
galaxies are selected by more complex criteria which are only indirectly
related to the host halo mass.  If we divide the total galaxy sample into 
different subsamples (e.g., by luminosity, color, light profile, ...), then 
the mass dependence shown in Figs.~\ref{bias_vs_M_res} and 
\ref{bias_vs_M_feat} will be observable as long as the
scatter in the mass-observable relation is not much larger than the
width of the features.  Fortunately, the latter typically corresponds to
a factor of 2 or more in mass, which should make these features 
detectable for a wide range of large-scale structure tracers.  Note
that the precise shape of $\Delta b_I(M)$ depends on the filter chosen
for the small-scale density field.  In principle, one could use a filter
matched to the Lagrangian profiles of dark-matter halos \cite{2010arXiv1010.2539D}.  
We have tried replacing the top-hat filter with a Gaussian, and found
only relatively minor differences, at the 7\% level for $\Delta b_I(M)$ (see Fig. \ref{tophat_vs_gaussian}). \\ 

\section{Conclusion}
We have shown that measurements of galaxy clustering could potentially be used as a probe of features in the inflationary potential. By computing the non-Gaussian correction to the halo bias, we revealed that features in the inflationary potential such as oscillations, bumps, or steps get imprinted onto the clustering properties of dark-matter halos.  While we have restricted ourselves to two generic models for which approximate forms of the oscillatory bispectrum are known, we expect this effect to be robust to the inclusion of more detailed bispectra. We note that this probe of primordial non-Gaussianity is complementary to CMB constraints as it probes very small scales where the microwave background becomes foreground dominated. On intermediate scales, the two approaches could be used in conjunction to cross correlate a possible feature in CMB data with a corresponding attribute in the clustering of dark-matter halos.

While showing the same scale dependence $\propto k^{-2}$, the predictions of 
the models considered here are strikingly different from the usually
considered local model.  In particular, they show significantly stronger 
effects for moderate halo masses ($10^{12}\!-\!10^{14}\:M_{\odot}/h$)
than the local model as compared to the effect at the high-mass end
($>\!10^{14}\:M_\odot/h$).  Thus, focusing on the most massive, highly biased
halos might not in general be the best way to design or optimize surveys for 
the search for primordial non-Gaussianity.  

The non-Gaussian models discussed here also make other predictions which are potentially observable with large-scale structure. While the study of these effects is beyond the scope of this paper, it would be interesting to correlate the non-Gaussian halo bias with features in the 
matter power spectrum and in the mass function of dark-matter halos.  The former could in principle be probed by weak lensing observations, and the latter through the abundance of galaxy clusters.  The bispectrum
of galaxies would also be a precise, albeit more complex and computationally
expensive, approach to testing these inflationary models.  The key advantage of the scale-dependent bias is,
however, that it is a unique signature of primordial non-Gaussianity which
is not easily mimicked by other effects.  We thus anticipate this observable
to be a robust probe of features in the inflaton potential.

\acknowledgments
We thank Marc Kamionkowski, Xingang Chen, and Kris Sigurdson for useful discussions. This work was made possible through a Michael Smith Foreign Study Supplement 
from the National Science and Engineering Research Council (NSERC) of Canada. FYCR wishes to thank the California Institute of Technology for hospitality while this work was completed.  FS was supported by the Gordon and Betty Moore Foundation at Caltech.


\end{document}